\documentclass[doublecol,amsmath,amssymb,figures]{epl2}

\newcommand{\BEQ}{\begin{equation}}     % Gleichungen Anfang ..
\newcommand{\BEA}{\begin{eqnarray}}
\newcommand{\EEQ}{\end{equation}}       % .. und Ende
\newcommand{\EEA}{\end{eqnarray}}
\newcommand{\eps}{\varepsilon}          % epsilon
\newcommand{\D}{{\rm d}}                % gerades d fuer Ableitungen
\newcommand{\II}{{\rm i}}               % gerades i fuer komplexe Einheit
\renewcommand{\vec}[1]{\boldsymbol{#1}} % Vektoren fettgedruckt

\usepackage{amsfonts}
\usepackage{amssymb}
\usepackage{amsbsy}

\title{Phase-ordering kinetics of two-dimensional disordered Ising models}
\shorttitle{Phase-ordering with disorder} 

\author{Florian Baumann\inst{1,2} \and Malte Henkel\inst{1} \and Michel Pleimling\inst{3}}
\shortauthor{F. Baumann, M. Henkel, M. Pleimling}

\institute{                    
  \inst{1} Laboratoire de Physique des Mat\'eriaux,\footnote{Laboratoire 
           associ\'e au CNRS UMR 7556}
  Universit\'e Henri Poincar\'e Nancy I, B.P. 239, \\
  F -- 54506 Vand{\oe}uvre l\`es Nancy Cedex, France \\
  \inst{2} Institut f\"ur Theoretische Physik I, 
  Universit\"at Erlangen-N\"urnberg, 
  Staudtstra{\ss}e 7B3, D -- 91058 Erlangen, Germany \\
  \inst{3} Department of Physics, Virginia Polytechnic
  Institute and State University, Blacksburg, VA 24061-0435,
  USA }

\pacs{05.70.Ln}{Nonequilibrium and irreversible
thermodynamics}
\pacs{64.60.Ht}{Dynamic critical phenomena}
\pacs{75.10.Nr}{Spin-glass and other random models}

\abstract{The phase-ordering kinetics of the ferromagnetic $2D$ Ising model 
with uniform disorder is
investigated by intensive Monte Carlo simulations. Taking into account
finite-time corrections to scaling, simple ageing behaviour is observed in
the two-time responses and correlators. The dynamical exponent $z$ and the 
form of the scaling functions only depend on the ratio $\eps/T$, where 
$\eps$ describes the width of the distribution of the
disorder. The agreement of the predictions of 
local scale-invariance generalised to $z \ne 2$ for the two-time 
scaling functions of response and correlations 
with the numerical data provides a direct test of 
generalised Galilei-invariance. 
}

\begin{document}

\maketitle

\section{Introduction}

A ferromagnetic 
system quenched from an initially disordered state into its coexistence phase 
with at least two equivalent equilibrium states undergoes phase-ordering
kinetics, driven by the surface tension between the ordered domains whose 
linear size grows as $L=L(t)\sim t^{1/z}$ where $z$ is the dynamical exponent. 
For a non-conserved order-parameter it is well-known that $z=2$, see 
\cite{Bray94} for a review. Phase-ordering is one of the instances where {\em 
physical ageing} occurs, by which we mean the following properties: (i) slow 
(i.e. non-exponential) dynamics, (ii) breaking of time-translation invariance, and
(iii) dynamical scaling. Because of the simple algebraic scaling of the linear
domain size $L(t)$, two-time correlation and response functions 
are expected to display the following simple scaling forms in the ageing regime 
(where the observation time $t$ and the waiting time $s$, that 
are both measured since the
quench, satisfy $t,s\gg t_{\rm micro}$ and $t-s\gg t_{\rm micro}$):
\BEA
C(t,s;\vec{r}) \hspace{-0.2cm} &:=& \hspace{-0.2cm} 
\langle \phi(t,\vec{r}) \phi(s,\vec{0})\rangle
\hspace{-0.05cm} = \hspace{-0.05cm} s^{-b}
f_C\left(\frac{t}{s},\frac{\vec{r}}{(t-s)^{1/z}}\right)~,~~~~\\
R(t,s;\vec{r}) \hspace{-0.2cm} &:=& \hspace{-0.2cm}
\left. \frac{\delta\langle \phi(t,\vec{r})\rangle}{\delta h(s,\vec{0})}
\right|_{h=0} \hspace{-0.2cm} = 
s^{-1-a} f_R\left(\frac{t}{s},\frac{\vec{r}}{(t-s)^{1/z}}
\right)~. \nonumber
\EEA
Here $\phi(t,\vec{r})$ is the space-time-dependent order-parameter,
whereas $h(s,\vec{r})$
is the conjugate magnetic field (spatial  translation-invariance will
be assumed throughout this paper) and $a$ and $b$ are ageing exponents. The
scaling functions $f_{C,R}(y,\vec{0})\sim y^{-\lambda_{C,R}/z}$ for $y\to\infty$
which defines the autocorrelation exponent $\lambda_C$ and the autoresponse
exponent $\lambda_R$. For phase-ordering kinetics, it is generally admitted that
$b=0$ and simple scaling arguments show that $a=1/z$. For an initial 
high-temperature state and for pure ferromagnets, $\lambda_C=\lambda_R$ 
is independent of the known equilibrium exponents 
\cite{Bray94,Godreche02,Henkel08}. More detailed information is contained in
the {\em form} of the scaling  functions $f_{C,R}(y,\vec{u})$. Indeed, for
the phase-ordering kinetics of pure ferromagnets where $z=2$, it has been shown
that dynamical scaling can be extended to a {\em local} scale-invariance (LSI)
which in particular implies co-variance of the linear responses under 
transformations $t\mapsto t/(\gamma t+\delta)$ in 
time. The other important ingredient is the Galilei-invariance of the 
deterministic part of the associated stochastic Langevin equation. 
Explicit predictions for the scaling functions $f_{C,R}$ follow and
numerous tests have been performed in a large variety of models and different
physical situations, see the recent
reviews \cite{Henkel07a,Henkel07b,Henkel07c} and references therein.

Here, we shall present tests of LSI in cases where
the dynamical exponent $z\ne 2$. The foundations of the theory were presented
some time ago in \cite{Henkel02} and have been  reformulated 
recently \cite{Baumann07a}. Several tests of LSI 
were performed in systems described by
{\em linear} Langevin equations in cases where 
either $z=4$ \cite{Roethlein06,Baumann07d} or else
$0<z<2$ \cite{Baumann07c}. In this letter, we present the first test of LSI
as reformulated in \cite{Baumann07a} with $z\ne 2$ in a model
where the corresponding Langevin equation is {\em non}-linear.

We consider a two-dimensional ferromagnetic Ising model with quenched disorder.
The nearest-neighbour hamiltonian is given by \cite{Paul04,Rieger05}
\BEQ \label{Ising_desord}
\mathcal{H} = -\sum_{(i,j)} J_{ij} \sigma_i \sigma_j, \quad
\sigma_i = \pm 1~.
\EEQ
The random variables $J_{ij}$ are uniformly distributed over
$[1-\eps/2,1+\eps/2]$ where $0 \leq \eps \leq 2 $. The model has
a second-order phase transition at a critical temperature $T_c(\eps)>0$ 
between a paramagnetic and a (diluted) ferromagnetic state. Using
heat-bath dynamics such that the order-parameter is non-conserved and starting
from a fully disordered initial state, phase-ordering occurs where the
dynamical exponent is given by 
\BEQ \label{z}
z = z(T,\epsilon) = 2 + \epsilon/T~.
\EEQ
This formula can be derived from phenomenological scaling arguments which assume
that the disorder is creating defects with logarithmically distributed barrier
heights parametrised by the constant $\epsilon$ 
\cite{Paul04,Paul05} and from field-theoretical studies in the Cardy-Ostlund 
model \cite{Schehr05}. Simulations of the linear domain size $L(t)\sim t^{1/z}$
\cite{Paul04} and of the scaling of the autoresponse function $R(t,s;\vec{0})$
\cite{Henkel06a} also confirm eq.~(\ref{z}) and furthermore suggest the empirical 
identification $\epsilon=\eps$. 

Therefore, since $z$ depends continuously on control parameters, the disordered
$2D$ Ising model offers a nice possibility to test universality 
and especially to test LSI for several values of $z$.

\section{LSI-predictions}

We now state the predictions of LSI with an arbitrary value of $z$ 
for the two-time responses and 
correlators, whose derivation will be presented elsewhere 
\cite{Baumann07a,Baumann07b}. First, the response function reads
\BEQ
\label{lsi_resp}
R(t,s; \vec{r}) = R(t,s)\,
\mathcal{F}^{(\alpha,\beta)}\left(
\frac{\vec{r}}{(t-s)^{1/z}}\right)
\EEQ
where $R(t,s)$ is the autoresponse function \cite{Henkel02}
\BEQ
\label{lsi_autorep}
R(t,s) =r_0 s^{-a-1}
\left(\frac{t}{s}\right)^{1+a' -
\lambda_R/z} \left(\frac{t}{s} -1 \right)^{-1-a'}
\EEQ
where $a'$ is a further exponent and the space-time part is 
\BEQ
\label{lsi_espacetemps}
\mathcal{F}^{(\alpha,\beta)}\left(
\vec{u} \right) = \int_{\mathbb{R}^d} \frac{\D
\vec{k}}{(2 \pi)^d} \, |\vec{k}|^{\beta} \exp
\left(\II \vec{u} \cdot \vec{k} - \alpha |\vec{k}|^z \right)~.
\EEQ
Here $\alpha$ is a dimensionful, non-universal parameter and $\beta$ an
universal exponent. Tests of eq.~(\ref{lsi_autorep}) check the
time-dependent symmetry $t\mapsto t/(\gamma t + \delta)$.  We already
presented such tests for the $2D$ disordered Ising model in detail 
\cite{Henkel06a,Henkel07a}. Here
we shall concentrate on a detailed test 
of the LSI-prediction eq.~(\ref{lsi_espacetemps}) for the space-time part  
in the disordered Ising model, which is the first time that Galilei-invariance 
generalised to $z\ne 2$ will be checked in a non-linear model. 

In practice, it is convenient to consider 
the integrated response function (thermoremanent magnetisation),
which is defined as
%%MH
%%MH ein Faktor h fuer die Magnetisierung
%%MH
\BEQ
M_{\rm TRM}(t,s; \vec{r}) := h \int_0^s \D \tau\, R(t,\tau; \vec{r})~.
\EEQ
In \cite{Henkel03}, we showed that already for $z=2$  
$M_{\rm TRM}(t,s;\vec{r})$
may contain important finite-time corrections to scaling. Adapting that procedure 
to an arbitrary value of $z$, we find 
\BEA \label{Mtrm}
& &M_{\rm TRM}(t,s; \vec{r}) = M_{\rm eq}(t-s; \vec{r})
+  r_0 M_{\rm age}\left( \frac{t}{s},\frac{\vec{r}}{s^{1/z}} \right)
%s^{-a}
%f_M\left(\frac{t}{s},\frac{\vec{r}}{s^{1/z}} \right)
\nonumber \\
& & \hspace{1.5cm} + r_1 t^{-\lambda_R/z}
\mathcal{F}^{(\alpha,\beta)}\left(\frac{\vec{r}}{t^{1/z}}
\right) 
\EEA
where $r_0,r_1$ are constants and $M_{\rm eq}(t-s;\vec{r})$ is the equilibrium 
contribution which vanishes for the chosen fully disordered initial conditions. 
The ageing part of the thermoremanent
magnetisation is given by
\BEQ
M_{\rm age}\left( \frac{t}{s},\frac{\vec{r}}{s^{1/z}} \right) = s^{-a} 
f_M\left(\frac{t}{s},\frac{\vec{r}}{s^{1/z}} \right)
\EEQ
\BEA
\label{di_fm}
& & \hspace{-1.0cm} f_M(y,\vec{w}) = \int_0^1 \D v \, (1-v)^{-1-a} \left(
\frac{y}{1-v}\right)^{1 + a' - \lambda_R/z} \nonumber \\ & &
\hspace{-0.5cm} \times \left(\frac{y}{1-v} -1
\right)^{-1-a'}\mathcal{F}^{(\alpha,\beta)}\left(\vec{w} \,
(y-1+v)^{-1/z}\right).
\EEA
As for the space-time response in the pure Ising/Potts model quenched to $T<T_c$ 
\cite{Henkel03,Lorenz07} and as 
for the autoresponse function in the disordered Ising model \cite{Henkel06a},
we find in many cases that sizeable corrections to scaling as 
described by the last term in
eq.~(\ref{Mtrm}) arise and need to be subtracted from the simulational raw data
in order to obtain a data collapse.

Second, we consider the autocorrelation function, 
generalising the approach for $z=2$ \cite{Henkel04,Henkel07c}. In our new
approach to LSI, generalised Galilei-invariance with $z\ne 2$ implies 
the existence of integrals of motion involving higher powers of the momenta.
It follows that the four-point response function needed for the calculation
of $C(t,s)$ factorises into terms ${\cal F}^{(\alpha,\beta)}$ and also depends
on an `initial' condition \cite{Baumann07a}. Our final prediction for the
two-time autocorrelator will depend on the choice for that initial condition. We 
shall consider here the following  two cases. \\ 
{\bf 1.} As suggested by direct and straightforward 
comparison with the lattice model, we use
a fully decorrelated initial state, with $C(0,0;\vec{r}) = \delta(\vec{r})$. 
We find  
\BEA \label{fc1}
&& \hspace{-1.0cm} f_C(y) = f_C(y,\vec{0}) \\
& &\hspace{-0.5cm}= c_1 y^{\lambda_C/z} (y-1)^{d/z -
2 \lambda_C/z + 2 \beta /z} (y+1)^{-d/z - 2 \beta/z}
\nonumber
\EEA
where we have already taken into account that $b = 0$. The
amplitude $c_1$ remains a fitting parameter, whereas 
$\beta$ is related to $\lambda_R$ and $\lambda_C$ via
$\beta=\lambda_C-\lambda_R$. \\
{\bf 2.} From a physical point of view, we should consider on what time
scale the scale-invariant ageing behaviour really sets in. When plotting
the autocorrelator $C(t,s)$ over against $\tau:=t-s$, the data converge rapidly
towards a plateau and only for time differences $\tau\sim s^{\zeta}$ (where
$0<\zeta<1$ is a cross-over exponent) and for $s$ sufficiently large, the
scaling behaviour sets in \cite{Zippold00}. This co\"{\i}ncides with
the rightmost end of the plateau \cite{Andreanov06} 
and also with the time-scale on which deviations
from equilibrium are seen in the fluctuation-dissipation ratio \cite{Zippold00}. 
We therefore require an estimate
for the space-dependent correlator $C(s+\tau,s;\vec{r})$, 
with $\tau\sim s^{\zeta}$ for $s\to\infty$ and $|\vec{r}|$ sufficiently large. 
Direct simulations show 
that in this space-time regime
$\ln C \sim \vec{r}^2$, see figure~\ref{abb.1}d, hence we may write
$C\sim \exp(-g(\tau,s) \vec{r}^2)$ where the function $g$ is fixed from matching
it with the expected scaling behaviour. From the above consideration 
scaling behaviour should set in for $\tau\sim s^{\zeta}$, 
hence we expect that $g(s^{\zeta},s)\sim s^{2/z}$ and we shall therefore 
assume a gaussian behaviour
\BEQ \label{gl:Cinit}
C(s+\tau,s;\vec{r}) \sim \exp(-\nu\,\vec{r}^2/s^{2/z})
\EEQ
where $\nu$ is a free parameter. This simple ansatz naturally generalises the  
equal-time correlator for phase-ordering in pure systems, with $z=2$ 
\cite{Bray94} and may be further justified by generalising the derivation
given in \cite{Ohta82} for $z=2$ to the case at hand. We then find 
\BEA \label{gl:fc}
& & f_C(y) = c_2 y^{\rho} (y-1)^{-\rho-\lambda_C/z + 2
\beta/z + d/z} \label{fc3} \\ 
& & \times \int_{\mathbb{R}^d}\frac{\D \vec{k}}{(2 \pi)^d} |\vec{k}|^{2 \beta} 
    \exp\left( -\alpha |\vec{k}|^z (y-1) 
   - \frac{\vec{k}^2}{4 \nu}\right) \nonumber.
\EEA
where $\rho$ is a new parameter (related to the scaling
dimensions of the fields involved) and $c_2$ a normalisation constant. 
Again, we have $\beta=\lambda_C-\lambda_R$.
In eq.~(\ref{gl:fc}), the initial correlator (\ref{gl:Cinit}) 
gives rise to the factor $\exp(-\mu \, \vec{k}^2)$ ($\mu$ is a constant) 
in the integral. 
Since in the application to the $2D$ disordered Ising model 
the dynamical exponent $z>2$, from the first exponential factor in the
integral it is clear that only the small-momentum 
behaviour of the initial correlator will appreciably 
contribute to the autocorrelator scaling function. 
The parameter $\rho$ is found by requiring  
that for $y \rightarrow 1$ the autocorrelation function
should be both nonvanishing and finite. This leads to 
\BEQ \label{gl:rho}
\rho = (2\beta + d -\lambda_C)/z
\EEQ
We shall see below from our numerical results that this relation 
indeed holds true in the $2D$ disordered Ising model 
(within the numerical error bars).

\section{Results}

%%------------------------------------------------------------------------------%%
\begin{figure*}
{\par\centering \resizebox*{0.8\textwidth}{!}
{\includegraphics{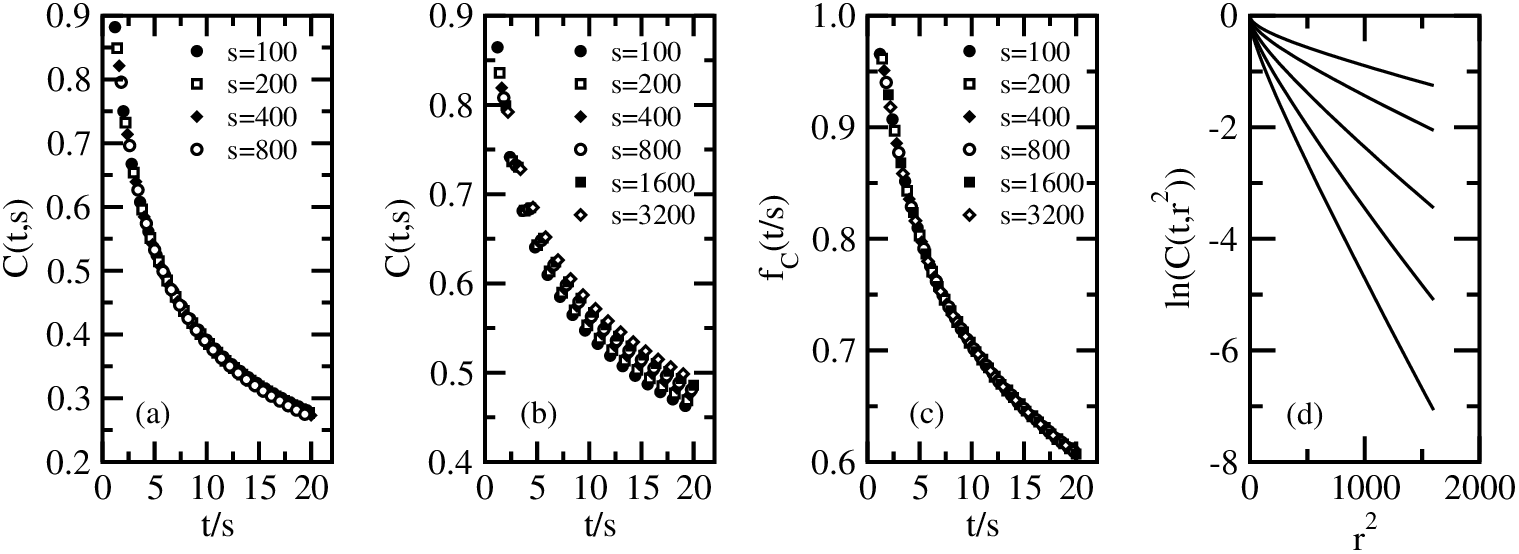}} \par}
\caption{
The autocorrelation as a function of $t/s$ for (a) $\eps=0.5$ and $T=1$, 
yielding $z=2.5$, and (b) $\eps=2$ and $T=1$, yielding $z=4$. 
Whereas in case (a) we find the
scaling behaviour of simple ageing, strong corrections to scaling behaviour
are seen in case (b). Identifying  the finite-time corrections to
scaling according to (\ref{ext_scaling}), 
we obtain $b' = 0.075$. Subtracting off the correction term, the scaling
behaviour of simple ageing is recovered, as shown in panel (c). In (d)
we show the equal-time space-dependent correlator $C(t,t;\vec{r})$ for
$\eps=0.5$ and $T=1.0$ for the times 
$t=[200, 300, 500, 1000, 2000]$ from bottom to top. 
\label{abb.1}
}
\end{figure*}
%%------------------------------------------------------------------------------%%

%%------------------------------------------------------------------------------%%
\begin{figure*}
{\par\centering \resizebox*{0.8\textwidth}{!}
{\includegraphics{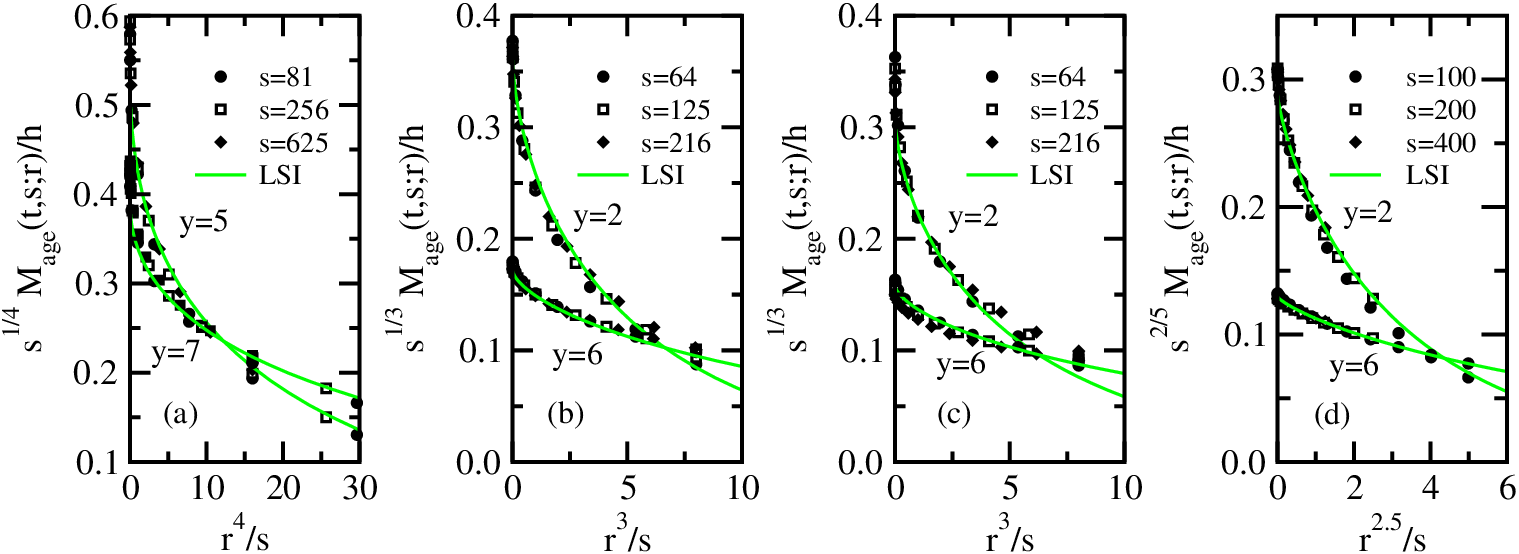}} \par}
%\onefigure[width=8.5cm]{Ising_des_espace_figure2.eps}
\caption[Mag]{
Scaling behaviour of the ageing part of the integrated spatio-temporal response 
$M_{\rm TRM}(t,s;\vec{r})$, for several waiting times $s$ and the
values (a) $\eps=2.0$, $T=1.0$, (b) $\eps=1.0$, $T=1.0$, (c) $\eps=0.5$, $T=0.5$
and (d) $\eps = 0.5$, $T = 1.0$,
as a function of $r^z/s$, where $z$ is given by (\ref{z}) and for two fixed
values of $y=t/s$. The full lines 
are the predictions of LSI, with parameters given in table~\ref{Tabelle1}. 
\label{abb.2}
}
\end{figure*}
%%------------------------------------------------------------------------------%%

%%------------------------------------------------------------------------------%%
\begin{figure*}
{\par\centering \resizebox*{0.8\textwidth}{!}
{\includegraphics{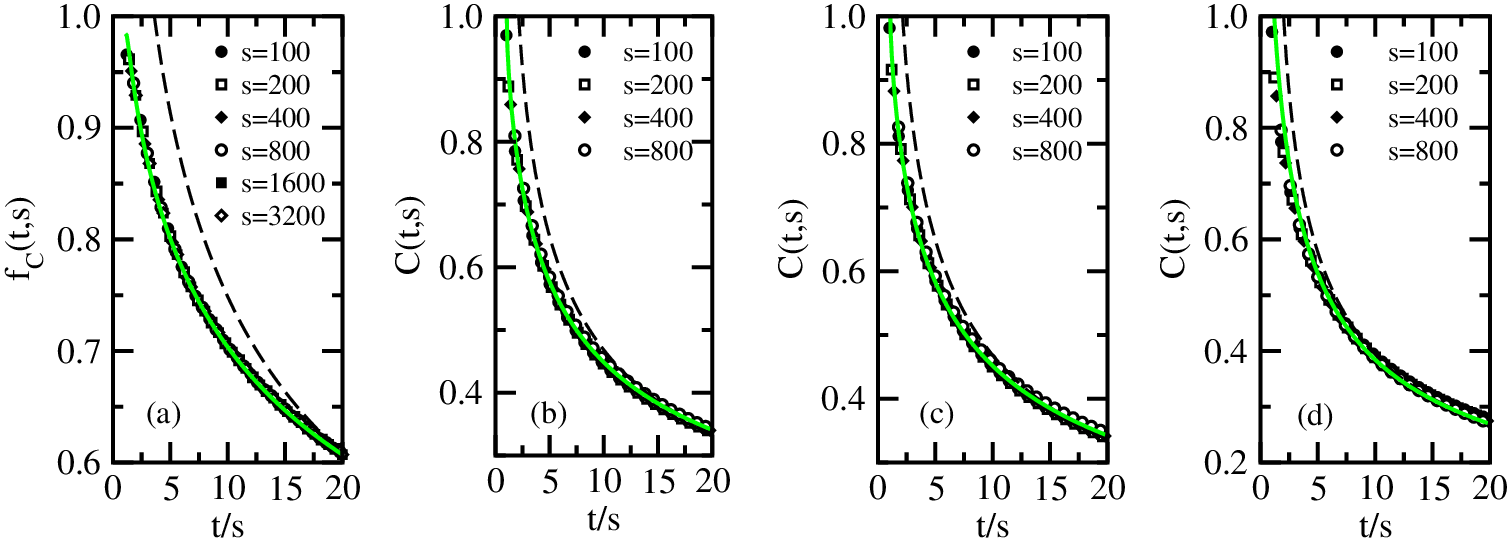}} \par}
%\onefigure[width=8.5cm]{Ising_des_espace_z=3_e1t1_n.eps}
%\onefigure[width=14.5cm]{Ising_des_espace_figure3_alt.eps}
\caption[Cor]{
Scaling behaviour of the autocorrelator $C(t,s)$, for the
values (a) $\eps=2.0$, $T=1.0$, (b) $\eps=1.0$, $T=1.0$, (c) $\eps=0.5$, $T=0.5$
and (d) $\eps = 0.5$, $T = 1.0$, as a function of $y=t/s$ and 
for several waiting times $s$. 
The dashed lines give the prediction (\ref{fc1}) with an
assumed totally disordered `initial' correlator and the full lines
give the LSI-prediction (\ref{fc3}), with the assumed long-ranged 
`initial' correlator (\ref{gl:Cinit}). The parameter values used are given in
table~\ref{Tabelle2}.
\label{abb.3}
}
\end{figure*}
%%------------------------------------------------------------------------------%%

We now compare these predictions with our numerical data. 
For the integrated response
we simulated systems with $300^2$ spins using the standard heat-bath algorithm. 
Prepared in an uncorrelated initial state 
corresponding to infinite temperatures,
the system is quenched to the final temperature in 
the presence of a random binary field
with strength $h = 0.05$. In order to obtain good 
statistics we averaged over typically
50000 different runs with different initial states 
and different realizations of the noise.
For the autocorrelation function we had to simulate 
systems containing $600^2$
spins in order to avoid the appearance of finite-size 
effects for the times accessed in the
simulations. It has already been noticed 
before \cite{Henkel06a,Henkel07a} that in
disordered magnets undergoing phase-ordering kinetics the 
autocorrelation may display 
finite-time corrections to scaling, leading to the extended scaling form
\begin{equation} \label{ext_scaling}
C(t,s) = f_C(t/s) - s^{-b'} g_C(t/s)
\end{equation}
with $b' > 0$. In fact, this extended scaling form is only needed for
dynamical exponents $z > 3$. We illustrate this in figure~\ref{abb.1}
for two different cases: $\eps=0.5$ and $T=1$, yielding $z=2.5$, in panel (a), 
and $\eps=2$ and $T=1$, yielding $z=4$, in panels (b) and (c). 
Whereas for $z=2.5$ one
readily observes the $t/s$-scaling behaviour of simple ageing, 
the existence of finite-time corrections to scaling is obvious for $z=4$. 
After subtracting off this correction, we recover the scaling of simple
ageing $C(t,s) = f_C(t/s)$, as shown in panel (c). We list in 
table~\ref{Tabelle2} the values of the exponent $b'$ of the 
correction term. It is natural that finite-time corrections should become
increasingly important with increasing values of $z$, since the domain size
$L(t)\sim t^{1/z}$ will grow more slowly for $z$ larger and scaling will set
in later.

\begin{figure}
\onefigure[width=8.5cm]{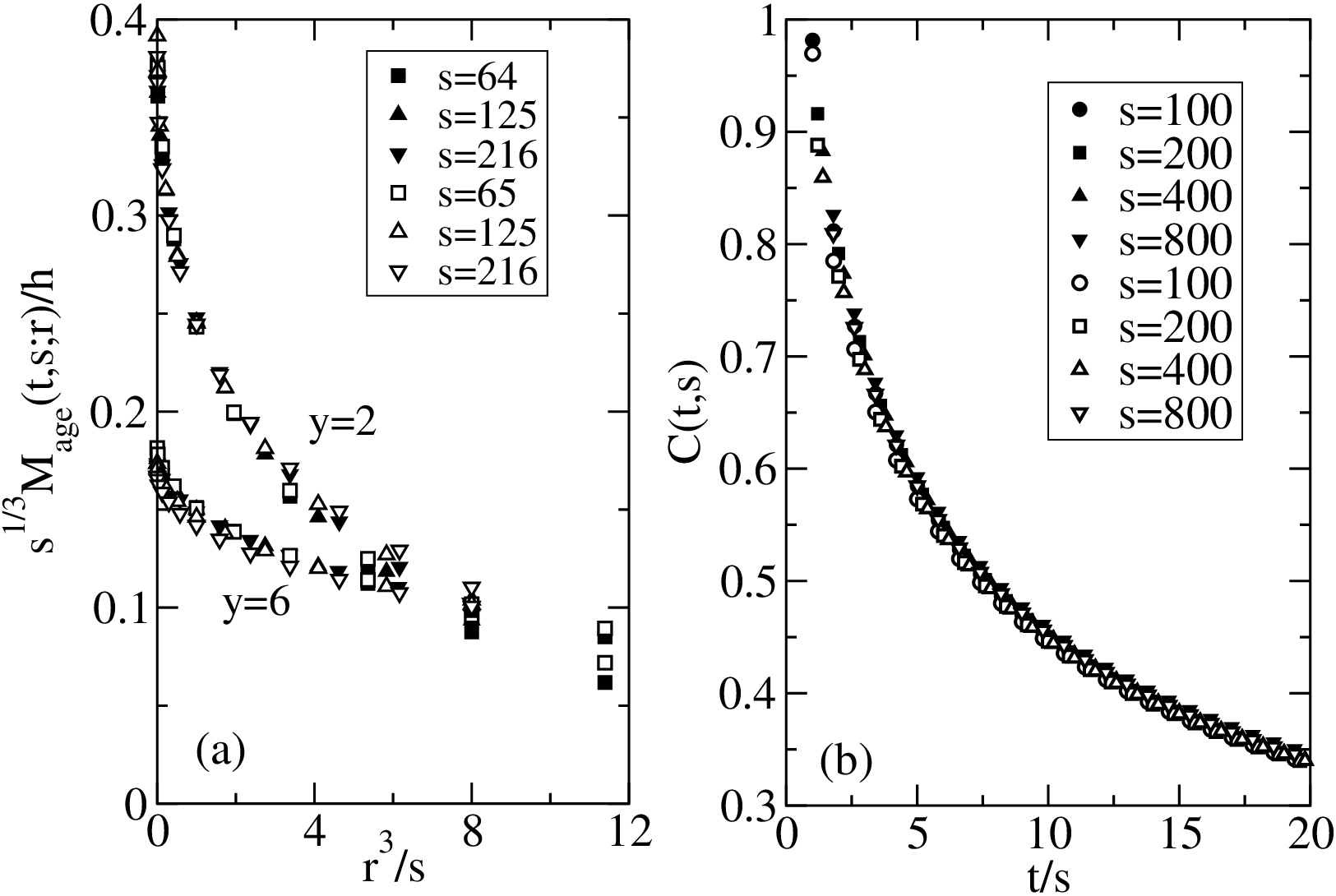}
%\onefigure[width=8.5cm]{Ising_des_espace_universalite.eps}
\caption[Comparacao]{Comparison of the scaling functions of the cases 
$\eps=1, T=1$ (filled symbols) and
$\eps=0.5, T=0.5$ (open symbols). 
In (a) the space-time
integrated response is shown for $y=2$ (upper curve) and $y=6$ (lower curve).
In (b) the autocorrelator is shown. The non-universal amplitudes have been
changed appropriately.
\label{abb.6}
}
\end{figure}

In figures~\ref{abb.2} and \ref{abb.3} 
we show the scaling functions
for the ageing part of the spatial 
thermoremanent magnetisation $M_{\rm age}(t,s;\vec{r})$ and the
autocorrelation function $C(t,s)$ for several combinations of $T$ and $\eps$. 
The observation of scaling of $M_{\rm age}$ as a function of $|\vec{r}|^{z}/s$
for $y=t/s$ fixed and with $z$ given by (\ref{z}) with $\epsilon=\eps$ 
gives a further confirmation for this dynamical exponent, 
see figure~\ref{abb.2}. 
We point out that the consideration
of the {\em spatio}-temporal response allows for a much more demanding test of
dynamical scaling than is possible by merely considering the autoresponse alone. 
Our results hence strengthen the conclusions of a simple power-law
scaling in the disordered Ising model reached 
earlier \cite{Paul04,Paul05,Henkel06a}. Our finding that the deviations from 
dynamical scaling seen in the raw data of the autocorrelation 
\cite{Henkel06a,Paul07,Henkel07a} can be explained in terms of a 
conventional finite-time correction makes it clear that more exotic proposals
such as `superageing' as proposed in \cite{Paul07} are not required. 
In any case, any evidence for `superageing' would have to argue against 
general theoretical arguments \cite{Kurchan02}, 
which use basic constraints from probability to assert that superageing is 
incompatible with scaling, before it could be considered to be conclusive.  
Furthermore, our data suggest a stronger universality in that not only
the dynamical exponent $z=2+\eps/T$ but also the {\em form} of the entire scaling
functions appear to depend merely on the ratio $\eps/T$ but not on these two
control parameters separately. We illustrate this in figure~\ref{abb.6},
where data for both the spatio-temporal response and the autocorrelator in 
the two cases $T=\eps=1$  
and $T=\eps=0.5$ are shown, which according to 
(\ref{z}) should give the same dynamical exponent $z=3$. Remarkably,
not only the ageing autocorrelation and autoresponse exponents indeed 
agree within the numerical precision of the data, 
but also the scaling functions are 
compatible with each other, up to normalisation (the non-universal amplitudes 
in figure~\ref{abb.6}a was changed appropriately). This is a strong indication
that the universality class of the non-equilibrium kinetics of the disordered
Ising model should not depend on $T$ and $\eps$ separately but rather only
on the dimensionless ratio $\eps/T$. Of course, this observation cannot be
explained in a `superageing' scenario that would be incompatible
with a simple scaling law $L(t)\sim t^{1/z}$ and with the well-accepted 
physical idea that in phase-ordering the domain size should be the only relevant
length scale \cite{Bray94}. Third, when considering the asymptotic
fall-off of the scaling functions, we find slightly negative values
of the exponent $\beta=\lambda_C- \lambda_R$, see table~\ref{Tabelle1}. 
More precise data would be needed to clarify whether
these results are significantly different from $\beta=0$, as one obtains
in pure systems \cite{Bray94,Picone04,Baumann07c}.

%%------------------------------------------------------------------------------%%
%%------------------------------------------------------------------------------%%

%%++++++++++++++++++++++++++++++++++++++++++++++++++++++++++++++++++++++++++++++%%
\begin{largetable}
\caption{Critical exponents and parameters of LSI for the
different values of $\eps$ and $T$ for the integrated
response function.}
\label{Tabelle1}\hspace{-0.6cm}
\begin{tabular}{|cc|cccclll|} \hline \hline
$\eps$ & $T$ & $z$ & $a = a'$ & $\lambda_R/z$ & $\alpha$ & \multicolumn{1}{c}{$\beta$} & \multicolumn{1}{c}{$r_0$} & \multicolumn{1}{c|}{$r_1$} \\ \hline
$0.5$& $1.0$ & $2.5$ & $0.40(3)$ & $0.61(1)$ & $0.24(2)$ &
$-0.10(5)$ &$0.0064(1)$ & $~~\,0.0025(2)$ \\ 
$0.5$ & $0.5$ & $3.0$ & $0.33(5)$ & $0.51(1)$ & $0.20(2)$ &
$-0.06(7)$ & $0.0049(1)$ & $~~\,0.005(2)$  \\ 
$1.0$ & $1.0$ & $3.0$ & $0.33(5)$ & $0.51(1)$ & $0.20(1)$ &
$-0.06(7)$& $0.00575(2)$ & $-0.02(1)$ \\ 
$2.0$ & $1.0$ & $4.0$ & $0.25(2)$ & $0.33(1) $ & $0.15(2)$ &
$-0.04(10) $& $0.0365(1)$ & $-0.035(2)$ \\ \hline \hline
\end{tabular}
\end{largetable}
%%++++++++++++++++++++++++++++++++++++++++++++++++++++++++++++++++++++++++++++++%%
\begin{largetable}
\caption{Critical exponents and parameters of LSI for the
different values of $\eps$ and $T$ for the autocorrelation
function.}
\label{Tabelle2}\hspace{-0.6cm}
\begin{tabular}{|cc|cc|c|cccc|c|} \hline \hline
$\eps$ & $T$ & $z$ & $\lambda_C/z$ & $c_1$  & $\rho$& $ (2 \beta + d -
\lambda_C)/z$ & $\nu$ &  $c_2$ &  $b'$\\ \hline
$0.5$& $1.0$ & $2.5$ & $0.570(5) $ & $1.51(2)$ & $0.14(2)$ & $0.16(1)$ & $0.31(2)$ &
$1.72(1) $ & --\\ 
$0.5$ & $0.5$ & $3.0$& $0.490(5)$ & $1.48(1)$ & $0.16(3)$&
$0.14(1)$ & $ 0.35(2)$ & $1.30(1) $ & --\\ 
$1.0$ & $1.0$ & $3.0$ & $0.490(5)$ & $1.48(1) $ & $0.16(3)$ & $0.14(1)$ & $0.36(2)$ &
$1.30(1)$ & -- \\ 
$2.0$ & $1.0$ & $4.0$ & $0.320(5)$
& $1.62(2)$ & $0.15(2)$ & $0.15(1)$ & $0.48(2)$ &
$1.12 (1)$ & $0.075$\\ \hline \hline
\end{tabular}
\end{largetable}

%%++++++++++++++++++++++++++++++++++++++++++++++++++++++++++++++++++++++++++++++%%
%\begin{largetable}
%\caption{ Critical exponents and parameters of LSI.C2 for the
%different values of $\eps$ and $T$.}
%\label{Tabelle2}\hspace{-0.6cm}
%\begin{tabular}{|cc|ccccccc|ccc|c|} \hline \hline
%$\eps$ & $T$ & $z$ & $a = a'$ & $\lambda_R/z$ & $\alpha$ & $\beta $ & $r_0$ & $r_1$ &  $\lambda_C/z$ & $c_0$ & $\kappa$ & $b'$\\ \hline
%$0.5$& $1$ & $2.5$ & $0.4(3)$ & $0.61(1)$ & $0.35(1)$ & $0.62(3)$ &$0.160(3)$ & $0.05(3)$ &  $0.570(5)$ & $1.51(2)$ &  -0.72(4) & --\\ 
%$0.5$ & $0.5$ & $3$ & $0.33(5)$ & $0.51(1)$ & $0.33(1)$ & $0.78(5)$ & $0.049(4)$ & $0.05(2)$  & $0.490(5)$ & $1.48(1)$ & -0.84(5) & --\\ 
%$1$ & $1$ & $3$ & $0.33(5)$ & $0.51(1)$ & $0.33(1)$ & $0.78(5)$& $0.14(3)$ & $-0.4(1)$ &  $0.490(5)$ &$1.48(1)$ & -0.84(5) & --\\ 
%$2$ & $1$ & $4$ & $0.25(2)$ & $0.33(1) $ & $0.15(2)$ & $0.56(10) $& $0.77(2)$ & $-0.7$ & $0.320(5)$ & $1.61(2)$& -0.60(8) & $0.075$\\ \hline \hline
%\end{tabular}
%\end{largetable}
%%++++++++++++++++++++++++++++++++++++++++++++++++++++++++++++++++++++++++++++++%%

After these preparations, we can now consider the {\em form} of the 
scaling functions and compare our data with the LSI-predictions
(\ref{di_fm}) and (\ref{fc1},\ref{fc3}), respectively. 
For the space-time response function, 
the four panels in figure~\ref{abb.2} clearly 
show an excellent agreement. Since (\ref{di_fm}) does not depend on the
initial correlations, any assumption about the initial correlations only
enters via the values of the parameters $\alpha,\beta$ which must the same as
for the autocorrelation. 
This is the first time that Galilei-invariance generalised to $z\ne 2$ 
\cite{Baumann07a,Baumann07c} could be directly confirmed
for a model with an underlying non-linear Langevin equation. It is 
non-trivial that a single theory is capable to reproduce the shapes of the
scaling functions for values of the dynamical exponents which vary considerably. 
Together with our previous test \cite{Henkel06a} of LSI for the autoresponse
function in this model, this is strong evidence that LSI captures the essence of
the dynamical scaling behaviour of the linear response. 

On the other hand, tests of LSI for the autocorrelation functions are conceptually
more difficult. First, we consider a fully disordered  initial state, as 
suggested by the na\"{\i}ve comparison with the lattice model. 
In figure~\ref{abb.3} we compare with the prediction
(\ref{fc1}), with parameter estimates listed in table~\ref{Tabelle2}. 
Clearly, one has at best a qualitative agreement and
particular for smaller values of $y=t/s$, the data deviate strongly from the
curve (\ref{fc1}). Should one take this as a {\it bona fide} indication that
LSI could not describe the ageing of the $2D$ disordered Ising model? 
We have already recalled above the general arguments \cite{Zippold00} which
suggest that ageing only sets in for time differences $\tau=t-s\sim s^{\zeta}$, 
with $s\gg t_{\rm micro}$ and $0<\zeta<1$. This implies that the form of the
scaling function should rather be related to the `initial' correlator
$C(s+\tau,s;\vec{r})$. In figure~\ref{abb.1}d we show that the leading 
space-dependent behaviour of the equal-time correlator $C(s,s;\vec{r})$ 
that strongly deviates from a totally uncorrelated correlator which means
that the system had time enough to build up strong 
spatially long-range correlations. We have also checked that the 
leading $\vec{r}$-dependence
is also recovered in space-time-dependent correlators $C(t,s;\vec{r})$, with
$y=t/s=2$ or $3$. Therefore, the assumed initial correlator eq.~(\ref{gl:Cinit})
appears to be in good agreement with direct numerical data for the asymptotic
behaviour of the `initial' correlator. That form is quite reminiscent
to the well-known analytic forms found in phase-ordering kinetics of pure 
systems (where $z=2$) \cite{Bray94} and is one of the most simple generalisations
to $z\ne 2$. We see from figure~\ref{abb.3} that the 
expression (\ref{fc3}) for the autocorrelator
$C(t,s)$ predicted by LSI agrees nicely with the data and also
confirm eq.~(\ref{gl:rho}). 
This is the first time that a consequence of the factorisation of the four-point
response function is confirmed in a system with a non-linear Langevin equation. 
Refining the chosen form (\ref{gl:Cinit}) of the initial correlator will merely 
lead to small corrections to (\ref{fc3}), in particular for $y\to 1$. 

\section{Summary} 
{\bf 1.} Taking into account strong finite-time corrections to scaling, 
we find simple ageing behaviour for both the two-time responses and
correlators of the $2D$ disordered Ising model. 
No trace of a 'superageing'  behaviour is seen. \\
{\bf 2.} Not only the  dynamical exponent  $z=z(T,\eps)$ as given by (\ref{z}),
but also the universal form of the scaling functions only depends on the 
ratio $\eps/T$. \\
{\bf 3.} The agreement of the space-time response predicted by LSI with the
numerical data constitutes the first direct confirmation of Galilei-invariance
generalised to $z\ne 2$ in a non-linear model of ageing. \\
{\bf 4.} The precise scaling form of the autocorrelator $C(t,s)$ does depend
on the initial space-time correlator at the onset of the ageing
regime. While the usual hypothesis of a fully uncorrelated order-parameter does
not describe the data, the asymptotic form eq.~(\ref{gl:Cinit})
leads to a good agreement 
between	LSI and the data. In this way, we have confirmed for the first time
the factorisation of the $2n$-point response functions, characteristic of 
LSI with $z\ne 2$ \cite{Baumann07a}, in a non-linear model of ageing.

The numerical results presented in this letter are strong evidence in favour of
the existence of a local scale-invariance of the `deterministic' part of the
stochastic Langevin equation which describes the phase-ordering kinetics of 
the $2D$ disordered Ising model. This is the first time that 
LSI could be confirmed for both correlators and responses 
in a non-linear model with a dynamical exponent $z\ne 2$. 
Further tests of LSI in different systems  with $z\ne 2$ would be welcome.

\acknowledgments
We thank P. Calabrese, B. Nienhuis and I.R. Pimentel for useful discussions. 
MH thanks the Centro de F\'{\i}sica T\'eorica e Computacional da Universidade 
de Lisboa (Portugal) for warm hospitality. 
We acknowledge the support by the Deutsche Forschungsgemeinschaft
through grant no. PL 323/2 and by the franco-german binational
programme PROCOPE. The simulations have been done
on Virginia Tech's System X.

%%################################################################################

\end{document}